# Fermi Pulsar Revolution

**Patrizia A. Caraveo**
*Affiliation IASF-INAF*
*Address, Via Bassini, 15 -20133 Milano ITALY*
*E-mail:* pat@iasf-milano.inaf.it

2009 has been an extraordinary year for gamma-ray pulsar astronomy and 2010 promises to be equally good. Not only have we registered an extraordinary increase in the number of pulsars detected in gamma rays, but we have also witnessed the birth of new sub-families: first of all, the radio-quiet gamma pulsars and later an ever growing number of millisecond pulsars, a real surprise.
We started with a sample of 7 gamma-ray emitting neutron stars (6 radio pulsars and Geminga) and now the Fermi-LAT harvest encompasses 24 "Geminga-like" new gamma-ray pulsars, a dozen millisecond pulsars and about thirty radio pulsars. Moreover, radio searches targeted to LAT unidentified sources yielded 18 new radio millisecond pulsars, several of which have been already detected also in gamma rays. Thus, currently the family of gamma-ray emitting neutron stars seems to be evenly divided between classical radio pulsars, millisecond pulsars and radio quiet neutron stars.

## 1. Introduction

With the demise of the CGRO mission in 2000, gamma-ray pulsar astronomy entered a long hiatus with the family of gamma-ray pulsars being frozen to 7 objects, namely the radio pulsars Crab, Vela, PSRB1055-52, PSRB1706-44, PSRB1951+32, PSRB1509-58 (only detected at low energy) and the radio quiet Geminga [1,2]. While radio observers discovered several new pulsars within the unidentified EGRET sources error boxes[e.g.3], the lack of contemporary gamma-ray data made the search for gamma pulsed signal a frustrating exercise, leading, at most, to tantalizing results [4].
Despite the small number of confirmed gamma-ray pulsars, they are one of just two classes of objects identified as gamma ray sources. With about 70 Unidentified low latitude EGRET sources, it was clear that pulsars were to play a major role in future, more sensitive, gamma-ray experiments [e.g.5]. To improve our ability to find gamma-ray pulsars deeply imbedded in the noisy galactic plane the EGRET lesson pointed to two major avenues
In view of the paramount importance of the availability of contemporary radio and gamma observations, monitoring programmes were arranged for hundreds of promising radio pulsars, selected on the basis of their rotational energy loss weighted by the distance square [6,7].
Meanwhile, the expectation of a significant contribution from radio-quiet, Geminga-like, gamma ray pulsars lead to the development of new codes tailored to the search for pulsation in the sparse photon harvest typical of gamma-ray observations [8].
Both avenues are now bringing fruits, with some unexpected findings leading to the Fermi pulsar Revolution

## 2. Results beyond expectations

A preview of the Fermi ability to spot gamma-ray pulsars without any outside help came soon after the launch of the mission, when PSR J0007+7303 was detected within the CTA 1 supernova remnant[9,10]. Actually the gamma-ray source was known since EGRET times and had been extensively (yet inconclusively) studied in X-rays [11].
A faint X-ray source surrounded by diffuse emission pointed to a neutron star embedded in a PWN, but the timing signature was missing, partly owing to the source faintness. Taking advantage of the X-ray source coordinates to barycentrize the Fermi Lat photon arrival times, the application of the newly designed codes devoted to blind searches yielded the source periodicity.
The pulsar timing parameters unveiled a radio quiet neutron star with a kinetic age of 10.000 y (comparable with the CTA 1 SNR age) and a rotational energy loss intermediate between Vela and Geminga. Such a quick discovery of the first LAT radio quiet pulsar heralded a new era on gamma-ray pulsar astronomy.
Applying the blind search techniques to all unidentified gamma-ray sources (most of which dated back to EGRET and even COS-B times) 15 more gamma ray pulsars were found, establishing the radio quiet pulsars as a major fraction of the pulsar family[12].

13 of the 16 newly found pulsars are indeed unidentified EGRET sources, many of which had been already studied at X-ray wavelengths thus providing promising candidates whose positions could be injected in the data analysis, establishing a new synergy between X and gamma-ray astronomies [13]
Meanwhile, the radio monitoring campaigns, which had requested a lot of organizational efforts, started to pay back yielding the detection in gamma rays of 16 new radio pulsars, bringing the radio pulsars grand total to 21 sources [14].
Indeed, the first detection of a new radio pulsar is due to Agile, a small mission of the Italian Space Agency, which was able to pinpoint PSR J 2021+3651 [15] and to detect later few more pulsars, most notably PSR B1509-58 [16] which stands out owing to its very soft spectrum.
Vela, by far the brightest pulsar, was investigated in detail [17], to study the evolution of the light curve as a function of energy, to align carefully the gamma-ray light curve and the radio one and to establish the pulsar spectral shape (Figure 1). By pointing to a power law with exponential cutoff (with cutoff energy of few GeV) as the best fitting spectral shape, the LAT data solved a long standing debate. They proved that the gamma-ray photons were indeed produced far from the star surface thus clearly supporting the outer gap models as opposed to the polar cap ones. Owing to the exceptional statistics available for Vela, as well as for few other pulsars such as Crab and Geminga, it is possible to perform phase-resolved spectral analysis, assessing how the power low index and the cutoff energy vary throughout the pulsar rotation [18,19,20].

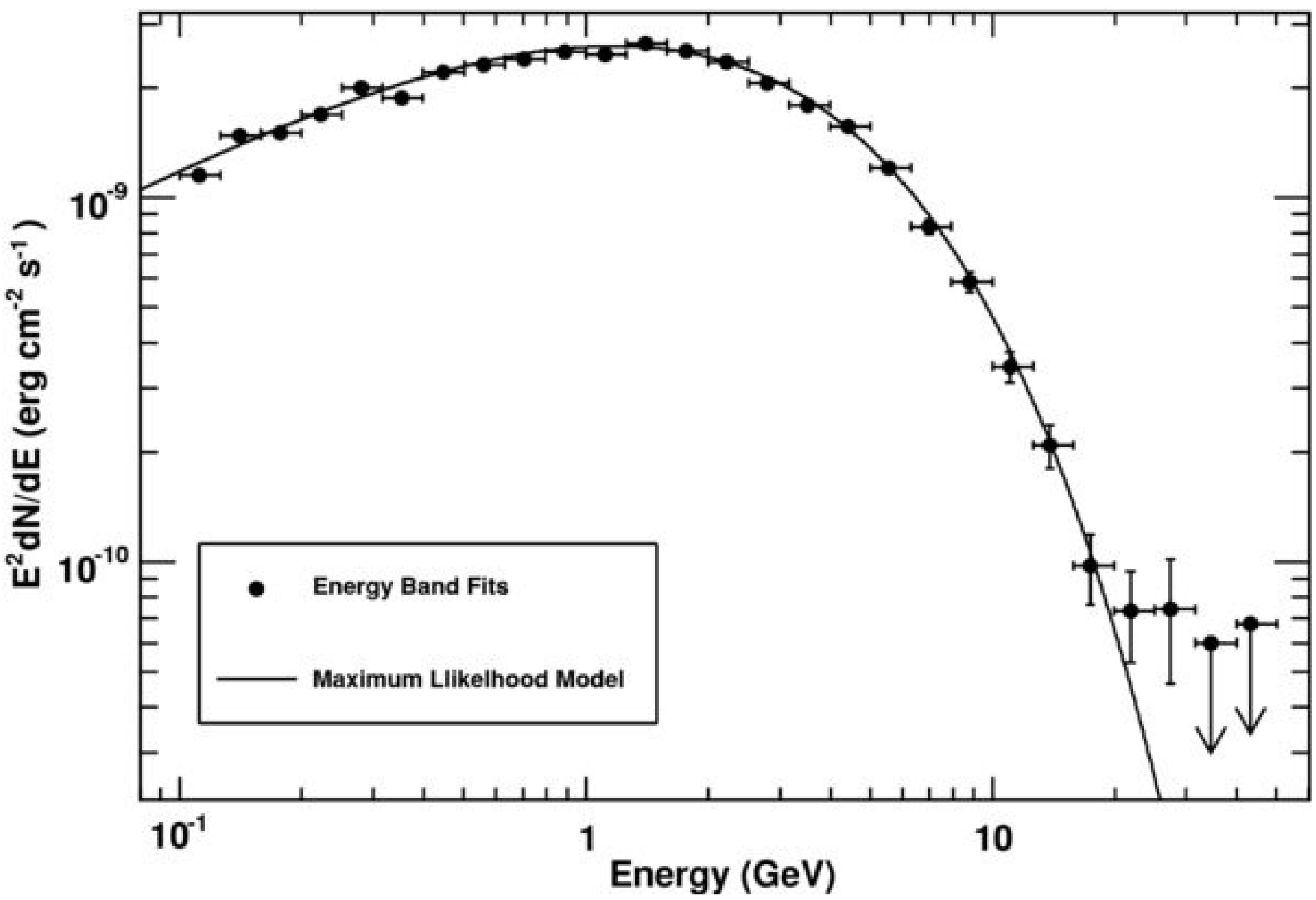

***Figure 1*** *Phase averaged spectrum of the Vela spectrum[18]. The data were fit assuming a power law with an exponential cutoff. The best fitted cutoff Energy is 1.36+/-0.15 GeV*

But another surprise was looming : msec pulsars were starting to be detected at gamma-ray energy. Although PSR 0218+4232 had been already tentatively detected in the EGRET data [21] msec pulsars were deemed to be generally unfit to produce high energy gamma-ray in view of their relatively low magnetic field.
However, the detection of 8 msec pulsars [22] changed completely the attitude forcing theoreticians to come up with suitable explanations linked e.g. to the strength of the magnetic field at the light cylinder (Figure 2).

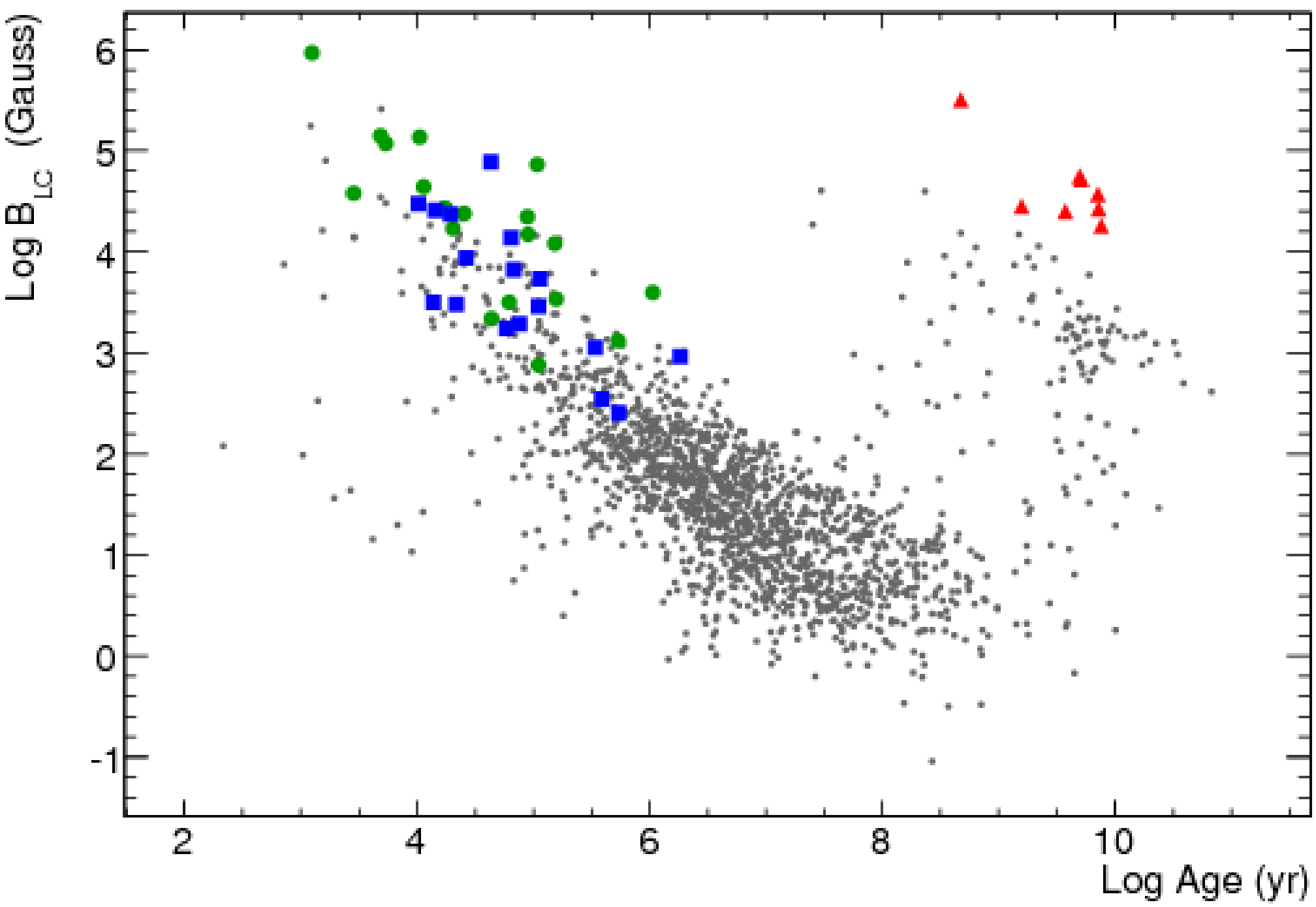

***Figure 2*** *Magnetic field strength at the light cylinder for the 46 pulsars in the first Pulsar Fermi catalog (from [14])****.*** *Green circles refer to radio pulsars, blue squares refer to gamma selected neutron stars, red triangles refer to msec pulsars*

Thus the first Fermi LAT pulsar catalogue went to press with a grand total of 46 pulsars 29 of which detected in radio (further divided between 8 msec and 21 "classical" pulsars) and 17 selected in gamma-rays (i.e. 16 discovered by LAT + Geminga). Indeed deep radio observations of the pulsars found through blind searches yielded detection for 3 of them, one rather normal looking while the other two exceedingly faint [23, 24].
The spectra of the 46 gamma-ray pulsars can all be fitted by power law with exponential cutoff. Their light curves are usually double peaked (with peak separation of 0.4-0.6) but a non negligible minority of single peaked pulsars do exists. With very few exceptions , the gamma ray peaks are not aligned with the radio ones.
To assess the pulsars gamma ray luminosity, we use a beaming factor of 1 for all pulsars. It is a big change wrt the EGRET time, when this parameter was $1/4\pi$, and is a direct consequence of the outer gap model. The gamma-ray light curves and spectral shapes point to high-altitude emission regions producing fan beams that cover a large fraction of the celestial sphere. Distance value is a major source of uncertainty for LAT pulsars, few of which have a measured parallax, while the majority should rely on dispersion measure or columnar absorption .
As shown in figure 3, the evolution of gamma-ray luminosity as a function of the pulsar rotational energy loss cannot be fitted by a single function . Even considering the distance uncertainties, a substantial scatter is present possibly arguing against the assumption of a common beaming factor for all objects. Msec pulsars seems to climb more steeply than normal and radio quiet pulsars which seems to evolve more gently. Such plot tell us more about our limitations than about the pulsar physics

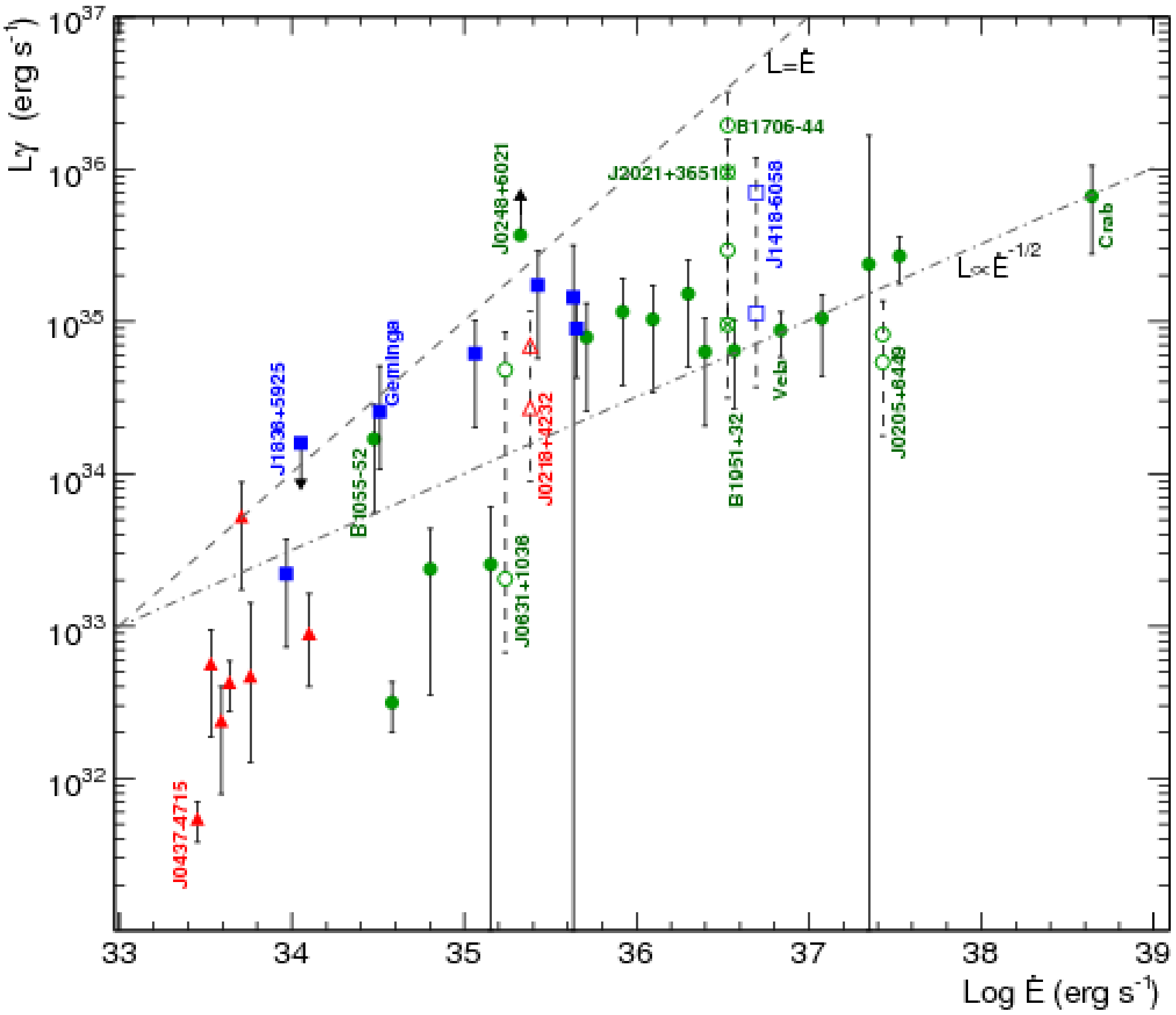

**Figure 3** *Gamma-ray luminosities versus rotational energy losses for the 46 pulsars in the first Fermi catalog (from [14])***.** *Green circles refer to radio pulsars, blue squares refer to gamma selected neutron stars, red triangles refer to msec pulsars. Pulsars with two distance estimates have two markers connected with a dashed line*
*The dot-dashed line and the dashed one are meant to guide the eyes showing the trend expected for a gamma-ray Luminosity proportional to the square root of the rotation energy loss or for a gamma-ray luminosity proportional to the rotational energy loss.*

After the completion of the first LAT pulsar catalogue, more pulsars were detected. 8 new objects emerged from the blind searches done on the sources of the first year Fermi Lat catalogue [25], while more classical pulsars as well as msec pulsar kept adding to the catalogue. The first Fermi catalogue list 56 pulsars[26] and the number is continuously growing.

### 2.1.1 One more breakthrough

A major breakthrough come via radio astronomy when targeted observations of unidentified LAT sources yielded the discovery of 18 field msec pulsars [27]. While this is a major discovery in its own, since it increases significantly the number of the known field msec pulsars in few months, it also point out that msec pulsars could play a major role in accounting for unidentified galactic gamma ray sources.

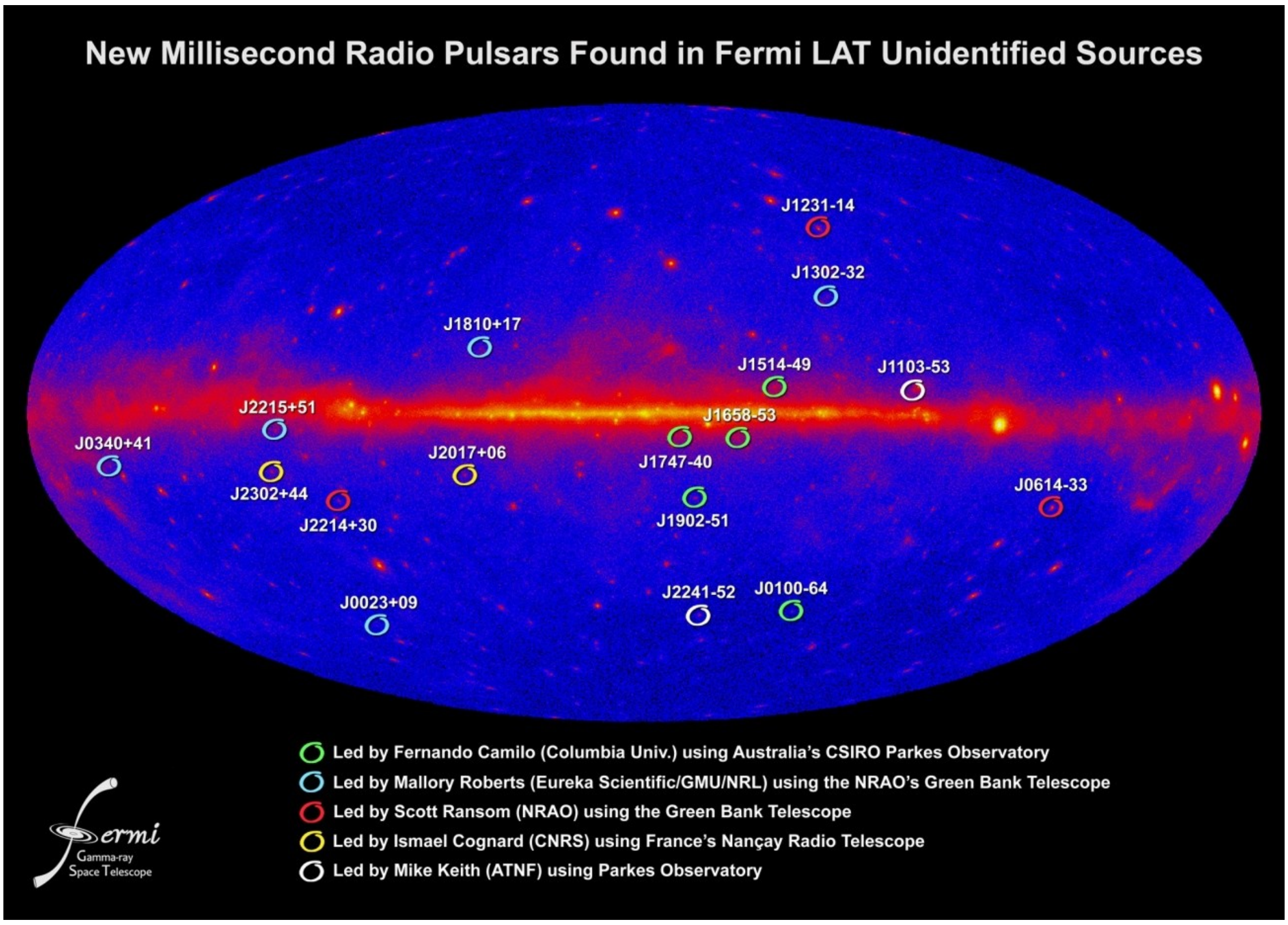

Figure 4 Msec pulsars detected while exploring error boxes of unidentified Fermi LAT sources.

## 3.Fermi revolution

In about 1 year gamma ray astronomy went from 7 gamma ray emitting neutron stars to 56 and counting, an astonishing jump rendered possible by the flawless Fermi LAT performances as well as by the extensive preparation work.
More importantly, new pulsar families have emerged claiming the undivided attention of the astrophysical community.
Apparently gamma-ray pulsars make up a class of sources “omnis divisa in partes tres” which classical radio loud objects, radio quiet (or radio faint) and msec ones claiming an equal share in source number.

The Fermi pulsar revolution having just began, such a harvest of results now needs to be digested and organized. Few questions call for an answer: are radio-quiet pulsars similar to radio-loud ones, just being viewed under different geometry? Now, for the first time we can try to address this question through multiwavelength observations. No matter what the answer will be, we will learn something on the pulsar emission mechanisms